\documentclass[twocolumn]{aastex7}

\usepackage{amsmath}
\usepackage{graphicx}
\usepackage{textcomp}
\usepackage{gensymb}
\usepackage{natbib}
\usepackage{longtable}
\usepackage{multirow}
\usepackage{float}
\usepackage{enumitem}
\usepackage{xspace}

\begin{document}

\def\rsun{R$_{\odot}$}
\def\msun{M$_{\odot}$}
\def\rstar{R$_{\star}$}
\def\mstar{M$_{\star}$}
\def\rearth{R$_{\earth}$}

\newcommand{\red}[1]{\textcolor{red}{#1}}

\newcommand{\Mstar}{\ensuremath{M_{\star}}\xspace}
\newcommand{\Mstariso}{\ensuremath{M_{\star,\mathrm{iso}}}\xspace}
\newcommand{\Rstar}{\ensuremath{R_{\star}}\xspace} 
\newcommand{\Rstariso}{\ensuremath{R_{\star,\mathrm{iso}}}\xspace} 
\newcommand{\Lstar}{\ensuremath{L_{\star}}\xspace} 
\newcommand{\rhostar}{\ensuremath{\rho_{\star}}\xspace}
\newcommand{\rhostariso}{\ensuremath{\rho_{\star,\mathrm{iso}}}\xspace}
\newcommand{\ageiso}{\ensuremath{\mathrm{age}_{\star,\mathrm{iso}}}\xspace}
\newcommand{\rhostarcirc}{\ensuremath{\rho_\mathrm{\star,circ}}\xspace}
\newcommand{\fe}{\ensuremath{\mathrm{[Fe/H]}}\xspace}
\newcommand{\femn}{\ensuremath{\left<\mathrm{[Fe/H]}\right>}\xspace}
\newcommand{\teff}{\ensuremath{T_{\mathrm{eff}}}\xspace}  
\newcommand{\logg}{\ensuremath{\log g}\xspace} 
\newcommand{\vsini}{\ensuremath{v \sin i}\xspace} 
\newcommand{\kepmag}{\ensuremath{m_\mathrm{Kep}}\xspace}
\newcommand{\kmag}{\ensuremath{m_\mathrm{K}}\xspace}
\newcommand{\Fbol}{\ensuremath{F_{\mathrm{bol}}}\xspace}
\newcommand{\Lbol}{\ensuremath{L_{\mathrm{bol}}}\xspace}
\newcommand{\LLbol}[1]{\ensuremath{L_{\mathrm{bol},#1}}\xspace}
\newcommand{\Mbol}{\ensuremath{M_{\mathrm{bol}}}\xspace}
\newcommand{\MMbol}[1]{\ensuremath{M_{\mathrm{bol},#1}}\xspace}

\newcommand{\Mp}{\ensuremath{M_p}\xspace} 
\newcommand{\Mcore}{\ensuremath{M_{c}}\xspace} 
\newcommand{\Menv}{\ensuremath{M_\mathrm{env}}\xspace} 
\newcommand{\Rp}{\ensuremath{R_p}\xspace}
\newcommand{\Rcore}{\ensuremath{R_c}\xspace} 
\newcommand{\fenv}{\ensuremath{f_\mathrm{env}}\xspace}
\newcommand{\Rpp}[1]{\ensuremath{R_{p,#1}}\xspace}
\newcommand{\PP}[1]{\ensuremath{P_{#1}}\xspace}
\newcommand{\teq}{\ensuremath{T_{\mathrm{eq}}}\xspace}
\newcommand{\Sinc}{\ensuremath{S_\mathrm{inc}}\xspace}
\newcommand{\Sincc}[1]{\ensuremath{S_\mathrm{inc,#1}}\xspace}
\newcommand{\Fxuv}{\ensuremath{\mathcal{F}_\mathrm{xuv}}\xspace}
\newcommand{\Tcirc}{\ensuremath{T_{\mathrm{max,circ}}}\xspace}
\newcommand{\tc}[2][red]{\textcolor{#1}{\emph{#2}}}

\newcommand{\ms}{m\, s$^{-1}$\xspace}
\newcommand{\kms}{\ensuremath{\mathrm{km\, s}^{-1}}\xspace}
\newcommand{\msyr}{m s$^{-1}$ yr$^{-1}$\xspace}
\newcommand{\Se}{\ensuremath{S_{\oplus}}\xspace}
\newcommand{\Me}{\ensuremath{M_{\oplus}}\xspace} 
\renewcommand{\Re}{\ensuremath{R_{\oplus}}\xspace} 
\newcommand{\um}{\ensuremath{\mu \mathrm{m}}\xspace}
\newcommand{\gcc}{g~cm$^{-3}$\xspace}
\newcommand{\Rsun}{\ensuremath{R_{\odot}}\xspace }
\newcommand{\Msun}{\ensuremath{M_{\odot}}\xspace}
\newcommand{\Mjup}{\ensuremath{M_{\mathrm{jup}}}\xspace}
\newcommand{\Rjup}{\ensuremath{R_{\mathrm{jup}}}\xspace}
\newcommand{\Lsun}{\ensuremath{L_{\odot}}\xspace} 
\newcommand{\K}{\ensuremath{\mathrm{K}}\xspace}
\newcommand{\dex}{\ensuremath{\mathrm{dex}}\xspace}
\newcommand{\dd}{\ensuremath{\mathrm{d}}\xspace}
\def\deg{\ensuremath{^{\circ}}}

\graphicspath{{newfigs/}}

\shorttitle{}
\shortauthors{Sullivan \& Gilbert}

\title{Planetary Architectures of Kepler Compact Multis with Binary Star Companions}

\author[0000-0001-6873-8501]{Kendall Sullivan}
\affiliation{Department of Astronomy and Astrophysics, University of California Santa Cruz, Santa Cruz CA 95064, USA}
\affiliation{Mullard Space Science Laboratory, University College London, Holmbury St Mary, Dorking, Surrey RH5 6NT, UK}
\email{kendall.sullivan@ucl.ac.uk}

\author[0000-0003-0742-1660]{Gregory J. Gilbert}
\affiliation{Department of Astronomy, California Institute of Technology, Pasadena, CA 91125, USA}
\email{ggilbert@caltech.edu}

\correspondingauthor{Kendall Sullivan}
\email{kendall.sullivan@ucl.ac.uk}

\begin{abstract}
Planets in binary-star systems exhibit demographic differences compared to planets in single-star systems. In particular, planets with binary-star hosts have a lower overall occurrence rate compared to their single-star counterparts, as well as a suppressed relative occurrence rate for sub-Neptunes ($R_p=2{-}4\Re$) compared to super-Earths ($R_p=1.0{-}1.5\Re$). These differences are most pronounced in close separation binaries ($\rho < 100$ au) which has been interpreted as a result of binary stars disrupting the protoplanetary disks of their stellar companions. The architectures of planetary systems --- i.e. the arrangements of planet sizes and orbits --- provide additional information about system formation and evolution. Architectures of single-star planetary systems are well studied, but architectures of binary-star planetary systems have not been investigated in detail. In this work, we analyzed a large sample of \textit{Kepler} planetary systems (162 planets in 118 binary-star systems; 880 planets in 544 single-star systems) to compare their architectures as a function of stellar multiplicity. We found that planets with binary-star hosts follow a similar ``peas-in-a-pod'' tendency toward uniformity in planet radii and log-uniformity in period spacing as planets with single-star hosts. However, we also detected modest ($2.5-3\sigma$) differences in period spacing and planet multiplicity, with binary-star systems having higher typical gap complexities (indicating more uneven spacing) and a higher prevalence of single planets. We interpret these results as evidence that binary stars primarily influence the planetary architectures of their stellar companions by shaping the protoplanetary disk at formation, with subsequent dynamical processing more gently altering the system architectures over secular timescales. 
\end{abstract}

\keywords{}

\section{Introduction}
Planets form concurrently with their host stars via oligarchic growth in the circumstellar disk \citep[e.g.,][and references therein]{Armitage2010}. Consequently, the properties of the protostar and disk environment (e.g., mass, metallicity, gas disk lifetime) are imprinted on the final arrangement of planetary masses, compositions, and orbits, called the system architecture. Studying present day exoplanetary system architectures therefore offers a powerful probe for exploring the relationships between system properties and the astrophysics of planet formation and evolution.

As an example, consider the observed abundance of ``peas-in-a-pod'' systems --- compact systems of small planets ($P \lesssim 100$ days, $R_p < 4 \Re$) with approximately equal sizes and approximately log-uniform spacing in period \citep{Millholland2017, Weiss2018}. The high prevalence of such systems suggests a paradigm of super-Earth and sub-Neptune formation as a global phenomenon that optimizes total energy \citep{Adams2020}. These planets may have formed \textit{in situ} close to their host stars from local material in the circumstellar disk \citep{HansenMurray2012, ChiangLaughlin2013}, or they may have formed further out followed by a period of resonant migration and dynamical disruption \citep{Izidoro2017}. The low but non-zero occurrence of planet pairs near mean motion resonance suggests that while some super-Earths and sub-Neptunes experience significant migration and capture into resonance, this is not the dominant channel of small planet formation \citep{LithwickWu2012, BatyginMorbidelli2013}. In contrast, hot Jupiter systems almost certainly formed by some entirely different mechanism that includes long-range migration from beyond the ice line \citep[e.g.,][and references therein]{DawsonJohnson2018}.

Such insights regarding the details of planet formation processes have largely been gleaned by studying planets orbiting single main-sequence stars \citep[e.g.,][]{WinnFabrycky2015}. Indeed, most of the over 6000 known planets fall into this category. And yet, approximately half of all solar-type stars in fact exist in binary or higher multiplicity star systems \citep[e.g.,][]{Duquennoy1991, Raghavan2010, Moe2017, Offner2023}. These binary-star systems offer a natural laboratory for studying planet formation, as the presence of an additional star in a system dramatically alters the radiation environment and mass reservoir available during planet formation. Unfortunately, accurately measuring the properties of planets in binary-star systems is challenging, and as a result whether and how stellar multiplicity alters planetary system architectures remains less thoroughly studied.

Recently, \citet{Sullivan2024} considered a sample of 286 confirmed and candidate planets from the \textit{Kepler} mission \citep{Borucki2010}, finding that for circumstellar planets in binary-star systems, the distribution of planetary radii depends significantly on binary separation. More specifically, they found that while wide binary stars ($\rho > 300$ au) and single-star systems host small planets with indistinguishable planetary radius distributions, close binary stars ($\rho < 100$ au) appear to host a suppressed population of sub-Neptunes. They concluded that this phenomenon likely arises from shortened disk lifetimes and/or diminished solid reservoirs in the close binary-star systems. Recent theoretical work supports this interpretation: by adapting the Bern model of planet formation \citep{Emsenhuber2021} to binary-star systems, \citet{Venturini2026} found that protoplanetary disk truncation is likely the dominant effect in altering the population of circumstellar (S-type) planets. We emphasize here that the \citet{Sullivan2024} results are also for circumstellar (S-type) planets in binary-star systems; circumbinary (P-type) planets \citep[see, e.g.,][]{Doyle2011, Armstrong2014, Martin2019} were not considered in that work, nor in the analysis presented in this paper.

Here, we investigate whether these previously observed differences in the properties of individual planets as a function of stellar binary separation also manifest as differences in system-level architectures. We consider the multiplicity (number of planets), size architecture (relative sizes of planets within a system), and spacing architecture (relative orbital period ratios within a system). We leave consideration of eccentricity and inclination architectures for future work.

This paper is organized as follows. In \S\ref{sec:sample} we describe our sample and briefly review the methods developed by \citet{Sullivan2022} to accurately characterize planets in binary-star systems. In \S\ref{sec:architecture} we describe the observed differences in architectures between single-star, close-binary, and wide-binary-star systems. In \S\ref{sec:discussion} we discuss implications for planet formation, and in \S\ref{sec:conclusion} summarize our main conclusions.

\section{Sample Selection}\label{sec:sample}

\begin{figure}
    \includegraphics[width=\linewidth]{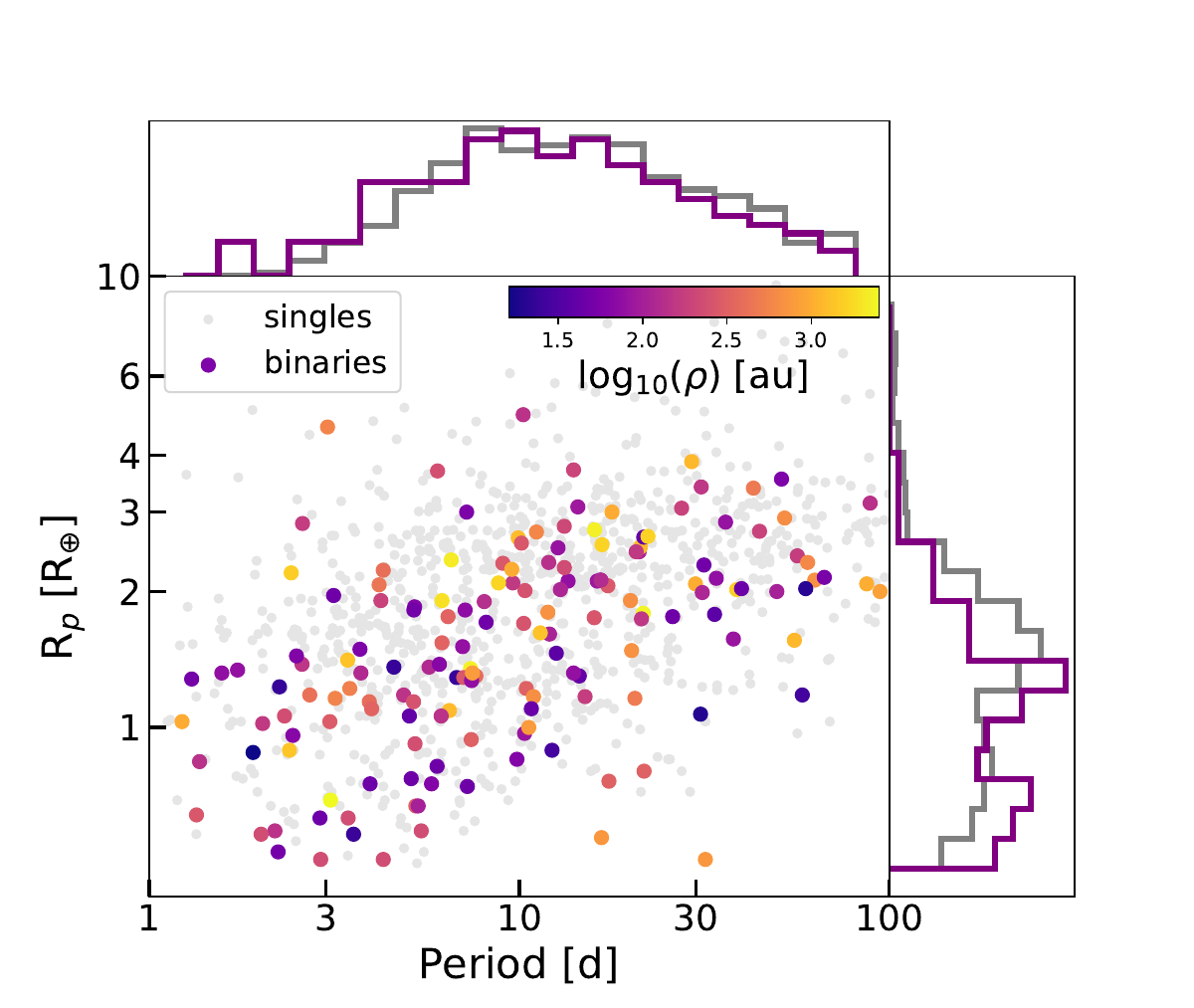}
    \caption{Period-radius plot for planets in single-star (gray points) and binary-star (colored points) systems considered in this work. Colors correspond to log$_{10}(\rho)$, i.e. the projected physical binary separation in au. We restricted the sample to planets with $1 < P < 100$ d and $0.5 < R_{p} < 4 \Re$. The planets in binary-star systems have a similar distribution in period-radius space as the planets in single-star systems, although the relative occurrence rate of sub-Neptunes ($2 < R_p < 4 \Re$) is suppressed in small-separation binary-star systems \citep{Sullivan2024}.}
    \label{fig:period_radius}
\end{figure}

\subsection{Planets in Binary Star Systems}\label{sec:binaries_sample}

\begin{figure}
    \includegraphics[width=0.9\linewidth]{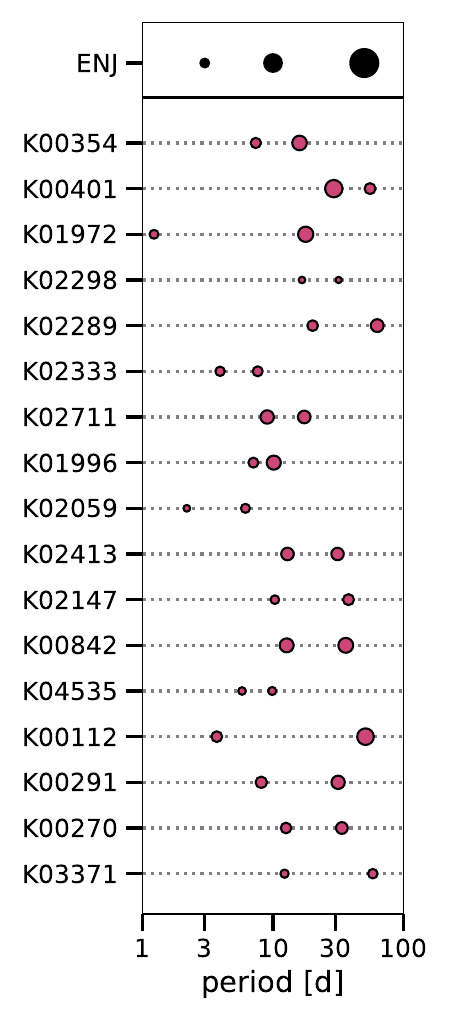}
    \caption{2-planet systems with binary-star hosts. Systems are ordered by binary separation, with close separations at the bottom and wide separations at the top. All $R_{p}$ assume the planets orbit the primary star. Earth, Neptune, and Jupiter radii (1, 4, and 10 \Re) are shown in the top row for comparison.}
    \label{fig:2pl_abacus}
\end{figure}

\begin{figure}
    \includegraphics[width=0.9\linewidth]{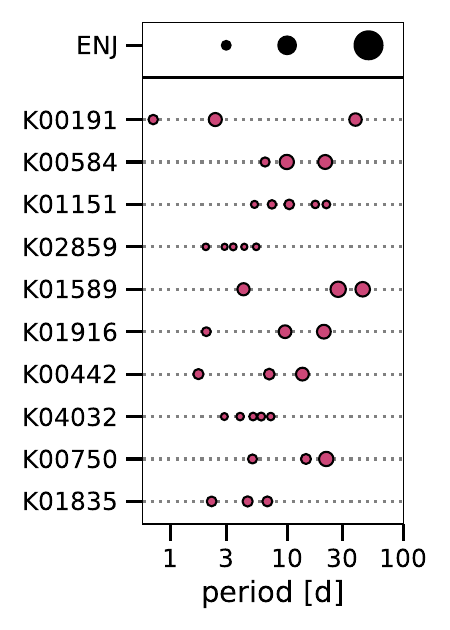}
    \caption{$3+$ planet systems with binary-star hosts. Systems are ordered by binary separations, with close separations at the bottom and wide separations at the top. All $R_{p}$ assume the planets orbit the primary star. Earth, Neptune, and Jupiter radii (1, 4, and 10 \Re) are shown in the top row for comparison.}
    \label{fig:multi_abacus}
\end{figure}

Our initial sample consists of 286 planets in 207 binary-star systems drawn from \citet{Sullivan2024}. These planets were all identified as candidate or confirmed planets by the \textit{Kepler} mission and had their stellar multiplicity confirmed from high-resolution imaging (speckle imaging, adaptive optics (AO) imaging, or both), using methods described in \citet{Sullivan2022}. 

Briefly, \citet{Sullivan2022, Sullivan2024} analyzed unresolved low-resolution optical spectra from the Hobby-Eberly Telescope, binary component relative brightnesses (companion contrasts) from AO and speckle imaging follow-up of the targets, and unresolved photometry from the Kepler Input Catalog (KIC; \citealt{Brown2011}). They used a two-component spectral model to simultaneously fit the data, and retrieved the properties of the individual stars in each binary. These properties were then used to adjust the previously measured radii for the planets in each system to correct the originally measured stellar radius and the flux dilution from the presence of the secondary star following \citet{Ciardi2015}. 

Following \citet{Sullivan2024}, we used Gaia DR3 \citep{Gaia2023} parallaxes to determine the distances. Where Gaia did not provide a parallax, we used the \citet{Mathur2017} distances. We calculated the physical binary separation using the distance and angular separations from \citet{Sullivan2024}. We used stellar properties ($R_\star, T_{\rm eff, \star})$ from \citet{Sullivan2024}. Finally, we added planet-to-star radius ratios $\Rp/R_\star$, impact parameters $b$ and periods $P$ from the final three \textit{Kepler} data releases, merging DR22 \citep{Mullally2015}, DR24 \citep{Coughlin2016}, and DR25 \citep{Thompson2018}, prioritizing the most recent values for each system.

To select the sample of planets in binary-star systems for this study, our goals were homogeneity and completeness. Thus, after removing known false positives, we implemented the following additional restrictions for data quality and reliability: impact parameter $b < 0.9$ (see \citealt{Petigura2020}, \citealt{Gilbert2022}), fractional radius error $\sigma(R_p)/R_{p} < 0.25$, period $P = 1-100$ days, planet radius $R_p = 0.5-16 \Re$, primary star effective temperature $T_{\rm eff, \star} = 4700 - 6500$ K, and primary star radius $R_{\star} = 0.7 - 1.4 \Rsun$. The stellar property restrictions were to ensure we were examining planets around main-sequence FGK stars. 

After filtering we retained 162 planets in 118 systems. Figure \ref{fig:period_radius} shows our sample of planets in binary-star systems superimposed over the background population of planets in single-star systems  (described in \S\ref{sec:singles_sample}). Figures \ref{fig:2pl_abacus} and \ref{fig:multi_abacus} show the architectures of the multi-planet systems with binary-star hosts, with Figure \ref{fig:2pl_abacus} showing all the two-planet systems and Figure \ref{fig:multi_abacus} showing the higher-order planet multiplicity systems. Planetary radii in Figures \ref{fig:period_radius}${-}$\ref{fig:multi_abacus} assume all planets orbit the primary star in the stellar binary, a choice we justify in \S\ref{sec:binaries_assignment}.

\subsection{Planets in Single-Star Systems}\label{sec:singles_sample}
To assemble a sample of comparison planets in single-star systems, we used the California-Kepler Survey (CKS) sample of FGK stars from \citet{Johnson2017} and their associated \textit{Kepler} planets. We imposed the same restrictions on the planet population as for the binary stars. To remove any remaining undetected binary stars, we imposed an additional constraint that the stellar renormalized unit weight error (RUWE) from a Gaia DR3 cross-match was RUWE $<$ 1.2 \citep{Gaia2016, Gaia2023}, to ensure that we were removing the majority of any remaining undetected binary-star systems from the CKS sample. Similarly, we removed any systems that had a planet radius correction factor PRCF $> 1.05$ in \citet{Furlan2017}. 

The presence of a nearby, dynamically impactful binary companion in the CKS sample would likely have been identified by high-resolution imaging follow-up \citep[e.g.,][]{Howell2011, Adams2012, Lillo-Box2012, Lillo-Box2014, Baranec2016, Kraus2016, Furlan2017}, excluding almost all small-separation binaries. In addition, one factor in determining whether a star was included in the KIC was whether it had a nearby stellar companion identified in the Kepler preliminary imaging \citep{Batalha2010}. Thus, any remaining binary companions in the CKS sample are likely low-mass, or at very wide separations. Low-mass companions and wide-separation companions both exert reduced influence on the primary star's protoplanetary disk \citep{Venturini2026}, so any lingering binary contamination in our single-star sample may be expected to have little impact on planetary system architectures.

Following our quality cuts, we retained 880 planets in 544 systems.

\subsection{Assigning Host Stars for Binary-Star Planets}\label{sec:binaries_assignment}

A complicating factor when considering the planetary architectures in binary-star systems is that we do not know with confidence which star the planets orbit or even whether all planets in a system orbit the same star. \citet{Sullivan2024} addressed this issue probabilistically, assigning each planet to the primary star with 80\% probability and to the secondary star with 20\% probability, then marginalizing over this uncertainty to derive the population-level distribution of $R_p$. When considering system architectures, absolute planet radius is less important, as we typically care more about planet sizes relative to their siblings. Whether planets in a binary-star system orbit the same or different star, however, can profoundly impact our understanding of the system multiplicity and spacing architecture.

Independently assigning planets to the primary or secondary star with the same 80/20 probability adopted by \citet{Sullivan2024} would assign nearly 1 in 3 two-planet binary-star systems a ``split'' architecture wherein one planet orbits the primary and the other orbits the secondary, and greater still for higher multiplicities. Given that (1) the typical projected spin-orbit alignment of solar-type binary stars is of order $\Delta i \sim 20^\circ$ \citep{JustesenAlbrecht2020}, (2) planet-planet mutual inclinations in single-star systems are quite small \citep[$\Delta i \sim 2^\circ$;][]{Fabrycky2014}, and (3) star-planet obliquities tend toward alignment \citep{Christian2022, Dupuy2022, Lester2023, Louden2024, Rice2024, Christian2025}, a ``split system'' detection rate over 1 in 3 is almost certainly too high. Although constraints on star-star and star-planet alignment remain imprecise, it is nevertheless reasonable to suppose that when multiple planets are detected in a binary-star system they orbit the same star in most cases. A similar line of reasoning was previously used to validate the low false-positive rate of \textit{Kepler} multis in general \citep{Lissauer2012}. Consequently, we assumed throughout our analysis that planets in high-multiplicity ($N > 1$) systems orbit the same stellar component, acknowledging that this simplification may introduce some level of bias. 

To investigate the impact of this assumption, we also performed our analysis assuming that 20\% of the planets orbited the secondary star rather than the primary star, matching the methodology of \citet{Sullivan2024}. The results of this analysis are shown in Appendix \ref{sec:appendix_secondary_mix}. To summarize, we found that the number of observed single-transiting systems increased while the number of multi-transiting systems decreased; population-level distributions of size and spacing architecture were largely unaltered. This result may be understood by recognizing that the most common outcome of ``splitting'' systems is to remove objects from the multiplanet population, thereby reducing statistical power but not substantially impacting observed multiplanet architectures. We also note that from visual inspection of Figures \ref{fig:2pl_abacus} and \ref{fig:multi_abacus} the architectures of the planetary systems with binary-star hosts are very similar to those in single-star systems, a coincidence which would be difficult to explain if ``split'' binary architectures are common. 

The one exception to the above assumptions was for orbital arrangements which were obviously unstable. We identified such systems by comparing period ratios of adjacent planets, flagging any systems in which $(P'/P)_{\text{multi}} < 1.15$, the minimum observed period ratio in single-star systems. Only 2 out of 43 binary-star systems were identified by this criterion: KOI-284 and KOI-1101. We removed these two systems from our analysis.

\subsection{Correcting for Binary Selection Effects}\label{sec:binaries_corrections}

\citet{Sullivan2026} found that \textit{Kepler}'s sensitivity to planets in binary-star systems drops dramatically once $R_{p} < 1.3 R_{\oplus}$. Thus, we also tested imposing a planet radius cut to both the binary-star and single-star samples to examine whether reduced detection sensitivity to small planets in the binary-star sample was biasing our results. We found that including or excluding small planets did not substantially change our results, but that the statistical power of the sample of planets in binary star systems was dramatically decreased, especially for the high planet multiplicity ($N \geq 3$) systems. Thus, we chose to proceed without implementing a planet radius cutoff at $R_{p} < 1.3 R_{\oplus}$. Plots demonstrating the results of the analysis with a planet radius restriction of $R_{p} > 1.3 R_{\oplus}$ are shown in Appendix \ref{sec:appendix_rp_cut}.

\section{System Architectures}\label{sec:architecture} 

After assembling our samples of planets in single- and binary-star systems, we calculated several diagnostic measures to compare the architectures of single-star vs binary-star systems. Whenever possible we split the sample of binary stars into ``wide binary'' and ``close binary'' samples (separations $>$ 300 au and $<$ 100 au, respectively). The close binary sample is expected to show the largest impact from stellar multiplicity \citep{Kraus2016, Moe2021, Sullivan2024}, but is relatively small (${\sim}40$ systems total) so it was often necessary to use the full binary sample marginalized over separation in order to avoid problems that arise from small-number statistics. Our analysis focused on the planet properties measured most precisely by transit surveys: planet multiplicity, relative planetary sizes, and orbital period.

\subsection{Planet Multiplicity}

\begin{figure}
    \includegraphics[width=\linewidth]{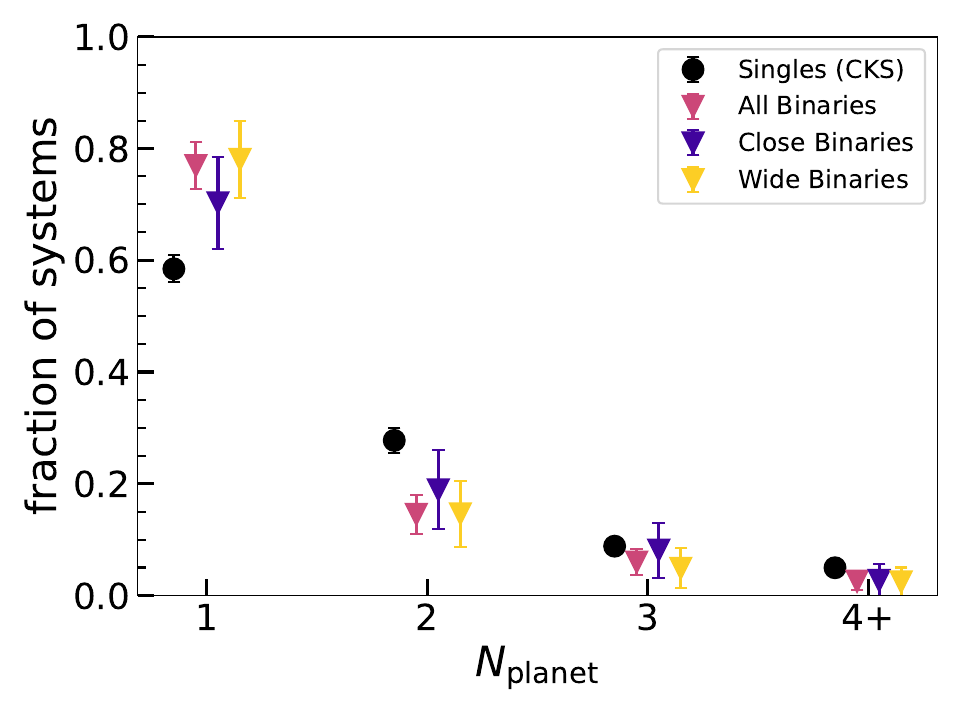}
    \caption{Fraction of planetary systems with 1, 2, 3, or 4+ planets detected for single stars (black circles), all binary stars (pink triangles), close binary stars (blue triangles) and wide binary stars (yellow triangles). The fraction of single-planet systems is larger for all binary-star planet host cases, and is statistically significant (3.0$\sigma$) based on a $\chi^{2}$ contingency test for all stellar binaries versus stellar singles.}
    \label{fig:multiplicity_fraction}
\end{figure}

Figure \ref{fig:multiplicity_fraction} shows the observed planet multiplicity fraction for planets orbiting single stars, all binary stars, close binary stars, and wide binary stars. In general, we found close agreement in the planet multiplicity distributions between the four groups, although single-planet systems appear to be slightly overrepresented in binary-star systems relative to single-star systems. To quantify this difference, we performed a chi-squared contingency test, finding $p = 0.003$ (${\sim}3.0 \sigma$) for the hypothesis that single-transiting systems are overrepresented in stellar binary stars compared to stellar single stars. The implications of this detection are discussed in \S\ref{sec:discussion}.

\subsection{Planet Size Architectures}

\begin{figure*}
    \includegraphics[width=\linewidth]{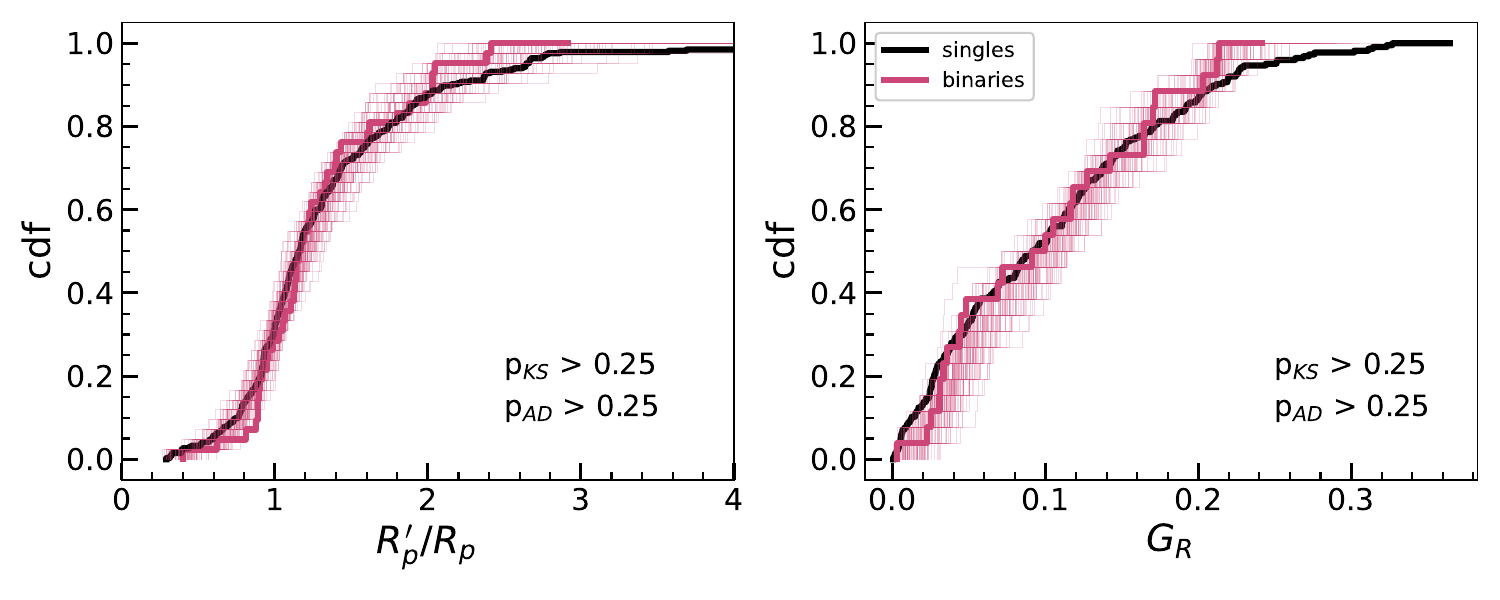}
    \caption{\textit{Left}: Cumulative distributions of the outer/inner radius ratios $R_p'/R_p$ for adjacent planet pairs in binary-star systems (pink) and single-star systems (black). \textit{Right}: Cumulative distributions of planet radius Gini indices $G_R$. The faint lines show the scatter of the distributions, calculated by drawing new $\Rp/\Rstar$ randomly from the normally distributed $\Rp/\Rstar$ errors and recalculating $R_p'/R_p$ and $G_R$; for visual clarity, we display these uncertainty bootstraps only for the binary-star systems and not the single-star systems. The binary-star systems and single-star systems are statistically indistinguishable within 1$\sigma$ for both $R_p'/R_p$ and $G_R$, suggesting that planets in binary-star systems have the same degree of size self-similarity as planets in single-star systems.}
    \label{fig:radius_ratios}
\end{figure*}

We next investigated planet size architecture in single-star vs binary-star systems by comparing radius ratios between adjacent planets in multi-planet systems. This analysis by definition could only be performed for systems with $N \geq 2$ planets, limiting our binary-star sample to 46 systems. Thus, we did not have sufficient numbers to divide our sample into wide and close binary sub-samples. 

The left panel of Figure \ref{fig:radius_ratios} shows the cumulative distributions (CDFs) of the radius ratios for single-star and binary-star systems. In both the single- and binary-star cases, the distribution peaks around 1, indicating that ``peas-in-a-pod'' self-similarity in radius is common for compact multis regardless of stellar multiplicity. To account for measurement uncertainty on planetary radii, we performed 100 bootstrap realizations by drawing radii randomly within the Gaussian errors for each planet; the solid lines show the distribution with the median planet radii. Kolmogorov-Smirnov (KS; \citealt{kolmogorov, smirnov1939}) and Anderson-Darling (AD; \citealt{anderson-darling}) statistical tests indicate that the two distributions are not statistically different, with all $\sigma < 1$. 

To provide another view of the radius similarity in each system, we calculated the system radius Gini index $G_R$ \citep{gini_index}. The Gini index is a measure of the homogeneity of a distribution, where a perfectly homogeneous system (in this context, a system with identical planet radii) would have a $G_R = 0$, while a perfectly inhomogeneous system (in this context, a system with a very small planet and a very large planet, for example) would have a $G_R$ close to 1. The Gini index is calculated as 0.5 times the sum of relative mean absolute differences:
\begin{equation}
G_{R} = \frac{\displaystyle \sum_{i = 1}^{N} \sum_{j = 1}^{N} |R_{i} - R_{j}|}{\displaystyle 2N\sum_{i=1}^{N}R_{i}}
\end{equation}
where $R_i$ is the radius of planet $i$ and $N$ is the total number of planets in each system. 

The right panel of Figure \ref{fig:radius_ratios} shows the CDF for the Gini indices for single-star and binary-star systems. As with the radius ratios, we accounted for radius measurement uncertainty via bootstrap trials. Based on KS and AD tests, $G_R$ is statistically indistinguishable between the single-star and binary-star systems. 

\subsection{Planet Spacing Architectures}

\subsubsection{Period Ratios}

To assess system spacing architecture, we calculated the outer/inner period ratios $P'/P$ of adjacent  planet pairs in each binary-star and single-star system with $N \geq 2$ planets. Orbital period can be measured directly and to high precision from a transit light curve, so bootstrap tests to include measurement uncertainty were unnecessary.

\begin{figure*}
    \includegraphics[width=\linewidth]{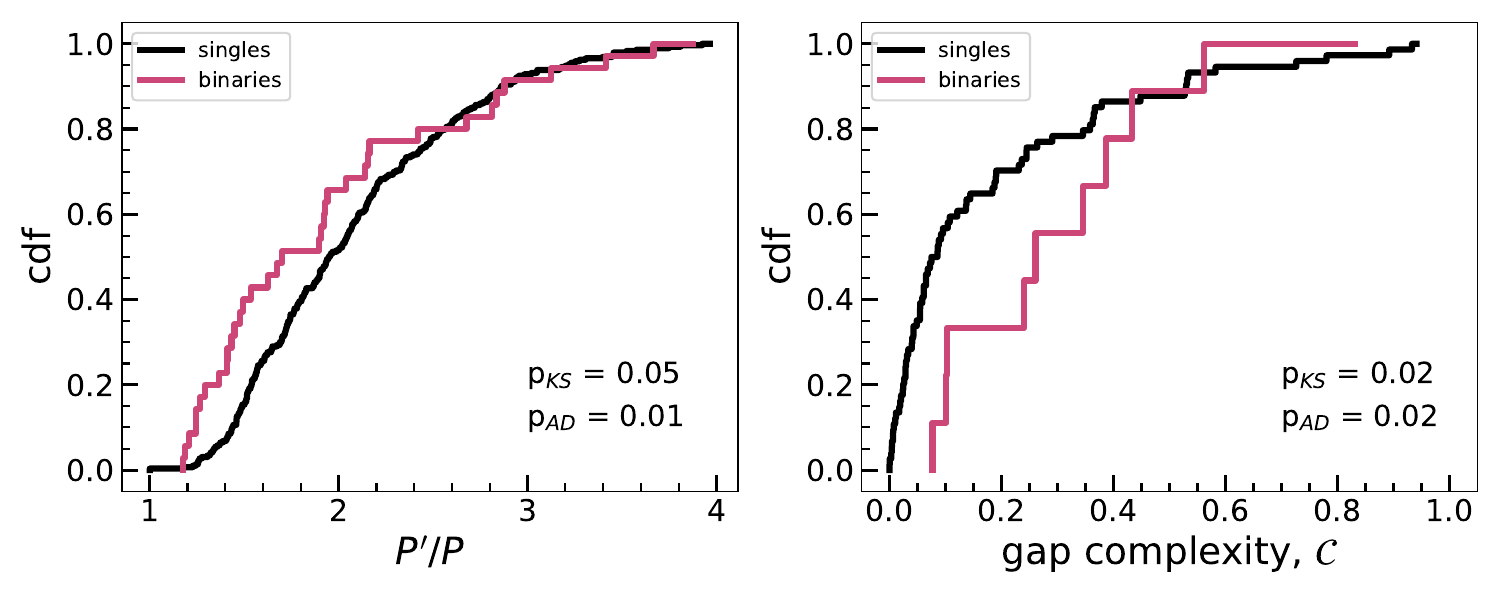}
    \caption{\textit{Left}: Cumulative distributions of the outer/inner period ratios $P'/P$ for adjacent pairs of planets in binary-star (pink) and single-star (black) systems. The two distributions have modestly significant differences, with $p_{KS} = 0.05\ (2.0\sigma)$ and $p_{AD} = 0.01\ (2.5\sigma)$ The planets in binary-star systems show a period ratio distribution that peaks at lower values than for those in single-star systems, suggesting that planets in binary-star systems are more tightly packed. The measurement errors on the planetary periods are negligible, leading to no apparent scatter in the CDF. \textit{Right}: Cumulative distributions of gap complexity $\mathcal{C}$ for planets in binary-star (pink) and single-star (black) systems. The distributions have modestly significant differences, with $p_{KS} = 0.02\ (2.4\sigma)$ and $p_{AD} = 0.02\ (2.4\sigma)$. The planets in binary-star systems tend toward higher gap complexity (i.e., less evenly spaced systems), which may indicate either intrinsically uneven spacing or higher mutual inclinations which lead to more non-transiting planets.}
    \label{fig:period_ratios_complexity}
\end{figure*}

The left panel of Figure \ref{fig:period_ratios_complexity} shows CDFs for $P'/P$. We found modestly statistically significant evidence for differences in $P'/P$ for binary-star and single-star systems, with $p_{KS} = 0.05\ (\sigma_{KS} = 2.0)$ and $p_{AD} = 0.01\ (\sigma_{AD} = 2.5$). The period ratios for the binary-star systems are preferentially closer to 1 than those for the single-star systems, suggesting that if the difference is real, then planets in binary-star systems are more tightly packed than those in single-star systems.

\subsubsection{Gap Complexity}

To further investigate the relationships between the periods for planets in each system, we calculated the gap complexity \citep{Gilbert2020}. Complexity is a statistical measure of the information content of a system, calculated by considering both the entropy and the (dis)equilibrium of a system (i.e., LMC complexity; \citealt{LMC1995}). The gap complexity is defined as:

\begin{equation}
    \mathcal{C} \equiv -K \left( \sum_{i = 1}^{n} p_{i}^{\star} \log_{10}(p_{i}^{\star}) \right) \left( \sum_{i = 1}^{n} \left(p_{i}^{\star} - \frac{1}{n}\right)^{2}\right),
\end{equation}

where 

\begin{equation}
    p_{i}^{\star} \equiv\frac{\log(P'/P)}{\log(P_{\rm max}/P_{\rm min})},
\end{equation}

$K$ is a normalization factor; $n$ is the number of gaps between planets in the system (i.e., $n = N-1$); and $P_{\rm min}$ and $P_{\rm max}$ are the global minimum and maximum periods in the system. $K$ is chosen such that $\mathcal{C}$ is always in the range (0, 1). A lower complexity implies more uniformity in a system: a system with planets evenly spaced in log-period will have $\mathcal{C} = 0$ while a highly disordered system will have $C\rightarrow1$.

The right panel of Figure \ref{fig:period_ratios_complexity} shows the CDFs of the gap complexities for the single-star and binary-star systems in the sample with at least $N \geq 3$ planets. Similarly to the period ratios, we found marginally statistically significant evidence for differences in the two populations, with  $p_{KS} = 0.02\ (\sigma_{KS} = 2.4)$ and $p_{AD} = 0.02\ (\sigma_{AD} = 2.4$). On average, the gap complexity for planets in binary-star systems is higher, suggesting less uniformly spaced planets.

\subsubsection{Mean-Motion Resonances}
To assess resonance behavior, we calculated the normalized distance from resonance $\Delta_k$ for both 1st- and 2nd-order mean-motion resonances \citep{Lithwick2012}:
\begin{equation}
    \Delta_k = \frac{P'}{P}\frac{j-k}{j} - 1,
\end{equation}
where $k=1$ for first-order resonances and $k=2$ for second-order resonances; $j$ is an integer such that $j > k$. We also calculated resonant neighborhood zeta statistic $\zeta_k$ \citep{Lissauer2011, Fabrycky2014}:
\begin{equation}
    \zeta_k = 3 \Bigg( \frac{k}{(P'/P) - 1} - \textbf{Round}\Big[\frac{k}{(P'/P) - 1}\Big]\Bigg),
\end{equation}
where as before $k=1$ for first-order resonances and $k=2$ for second-order resonances. The zeta statistic is scaled to run from -1 to +1 in the neighborhood of each resonance; the distribution of $\zeta$ relative to a test distribution may be used to identify statistically significant ``pile-ups'' of planet pairs near or far from resonance.

We found no statistically significant differences between the population of planets in binary-star versus those in single-star systems based on KS and AD tests on $\Delta_1$, $\Delta_2$, $\zeta_1$, and $\zeta_2$. Finding no obvious differences, we defer further discussion of resonances for future work where we can perform a more detailed analysis with explicit considerations for eccentricity, orbital stability, and dynamics.

\section{Discussion}\label{sec:discussion} 

\subsection{Single-Star vs. Binary-Star Systems}

Our analysis revealed two moderately significant (${\sim}2-3\sigma$) differences between single-star and binary-star system architectures: (1) single-transiting systems are more common for binary stars ($77\pm 4$\%) compared to single stars ($58\pm 2$\%), and (2) binary-star systems are more tightly packed than single-star systems, but also less evenly spaced. We found no statistically significant differences in the size architecture or resonance behavior of single-star versus binary-star systems.

Overall, the planetary architectures of single-star and binary-star systems are remarkably similar. For both populations, detected planet multiplicity drops monotonically from $N=1$ to $N=4+$, as expected from observational selection effects which make detection of multiple transiting planets less likely \citep[e.g.,][]{Winn2010, ZinkHansen2019}. The ``peas-in-a-pod'' trend toward uniform planet sizes and uniform spacing in log-period \citep{Millholland2017, Weiss2018} is also apparent in both the single-star and binary-star systems. It therefore appears likely that the same processes which dominate formation of compact multis in single-star systems also dominate in binary-star systems.

Before drawing any strong conclusions, we must rule out detection biases as the source of the observed trends. It is more difficult to detect planets in binary-star systems compared to single-star systems because the presence of a second star in the photometric aperture introduces flux dilution which reduces the apparent transit depth and limits the minimum size planet that can be detected \citep{Ciardi2015, Furlan2017, Lester2021, Sullivan2026}. How flux dilution alters planetary system architectures depends on whether binary-star planets adhere to the ``peas-in-a-pod'' trend. On the one hand, if binary-star planets are indeed ``peas-in-a-pod'', then flux dilution should affect all planets in a system roughly equally and therefore should not strongly influence the observed distributions of $G_R$ or $\mathcal{C}$. On the other hand, if binary-star planets are not ``peas-in-a-pod'', then small planets will be preferentially missed and the distributions of $N_{\rm planet}$ and $G_R$ will be pushed toward lower values (because the detectable dynamic range of $R_p$ is reduced) while the the distribution of $\mathcal{C}$ will be pushed toward higher values.

We observed $N_{\rm planet}$ and $\mathcal{C}$ trends toward lower and higher values, respectively, suggesting that small planets may have been missed in the binary star systems. However, we also found that the distribution of $G_R$ is indistinguishable for single-stars vs binary-stars. Although flux dilution is expected to push $G_R$ toward lower values (and $R_p'/R_p$ toward unity), a high degree of fine tuning would be needed to mimic the ``peas-in-a-pod'' phenomenon via detection bias alone. Indeed, this question has been thoroughly investigated in the literature for single-star systems \citep[see, e.g.,][]{Weiss2018, He2019, Wei2020, MurchikovaTremaine2020, Gilbert2020, WeissPetigura2020, Emsenhuber2021}, eventually reaching a consensus that the ``peas-in-a-pod'' phenomenon is unlikely to be an observational artifact. Thus, it seems unlikely that flux dilution is the sole driver of observed differences between single-star and binary-star planetary systems.

To more rigorously account for the influence of flux dilution, we repeated our analysis from \S\ref{sec:architecture} after applying a planet radius cutoff $R_p > 1.3$ to both single-star and binary-star systems (see \citealt{Sullivan2024, Sullivan2026} for discussion and justification of this threshold; see Appendix \ref{sec:appendix_rp_cut} for figures and detailed discussion). In this case, we found that the distribution of $G_R$ for binary-star systems now tended toward lower values compared to the distribution of $G_R$ for single-star systems. Conversely, the distribution of $P'/P$ for binary-star systems was now statistically identical to that of single-star systems. The number $N=2$ systems was reduced from 17 to 12, and the number of $N\geq3$ systems was reduced from 10 to 2; often entire systems were eliminated from our sample, a natural consequence of the intrinsic ``peas-in-a-pod'' architectures. Although applying the $R_p > 1.3 \Re$ cutoff did strengthen the trend toward uniform sizes and spacings in binary-star systems, the overall effect was minimal. We conclude that flux dilution alone is unlikely to have produced the observed trends in the data, and that we are observing the real astrophysical impacts of the companion star on the planetary system architectures.

The higher prevalence of single-planet systems and higher gap complexity for binary stars compared to single stars point toward increased dynamical excitation when a second star is present in the system. At the same time, the lack of a difference in size ratios and modest differences in period ratios suggest that any additional dynamical excitation induced by the binary companion is limited in scope. The tentative trend toward tighter spacing in binary star systems could indicate a survivor bias effect (i.e., tightly packed systems are more likely to persist against the gravitational influence of binary companion), a shepherding of planets inward by the binary companion, or minimal changes to the original system architecture after formation, with architectures primarily dictated by the original protoplanetary disk properties. Taken together, the overall picture is one in which the system architecture is mostly established by the time the protoplanetary disk dissipates, after which the companion star gently modifies the host star's planetary system architecture but does not outright destroy it.

The above interpretation is consistent with existing theoretical and empirical understanding of binary star influence on planet formation and evolution. Binary star companions are known to disrupt protoplanetary disks \citep[e.g.,][]{Heppenheimer1978, Artymowicz1994, Paardekooper2008, JangCondell2015, Zagaria2021}, increasing planet eccentricities and dynamically exciting their systems. Protoplanetary disks in binary systems are also smaller \citep{Harris2012, Zurlo2021} and shorter-lived \citep{Cieza2009, Kraus2012}, likely inhibiting planet formation in the outer disk and disrupting migration inward to the observed orbital periods of the mature planets. The smaller protoplanetary disks for each component of a binary star system might lead to lower average planet multiplicity (because less disk material is available to build planets) or might lead to the surviving multis being more tightly packed (because the spatial extent of the disks is smaller). 

After the protoplanetary disk dissipates, the dynamical influence of the secondary star is expected to continue altering system architectures. For example, binary star companions can induce migration to smaller orbital periods \citep{Wu2003, Takeda2009}, which sculpts system architectures to be more compact.

Induced migration might be classified as a relatively ``gentle'' dynamical process, by which systems are altered but the number of planets formed remains intact into system maturity (although we note that these mechanisms do not necessarily produce small-amplitude changes and could still be important for shaping the the system architectures). There are also some less gentle dynamical processes that can be driven by a binary star companion. In the most extreme cases, dynamical excitation induced by the binary companion could remove most or all of the planets from a system via collisions or ejections, producing a population of intrinsic single-planet systems. 

In less extreme cases, only a subset of planets would be removed, and the ``missing'' planets would leave behind a signature of more uneven period spacing and higher mutual inclinations and eccentricities. The broad similarity between single-star planetary architectures and binary-star planetary architectures suggests that the incidence of violent dynamics spurred by stellar binaries may be less important than more gentle dynamical processes. In the future, the relative importance of these various processes could be distinguished with follow-up Doppler campaigns to detect any surviving non-transiting planets and measure planet masses. Future work will measure eccentricities from the transit light curves and explore dynamical stability of the compact multi systems.

\subsection{Hints of Binary-Star Separation Dependence}
\begin{figure}
    \centering
    \includegraphics[width=\linewidth]{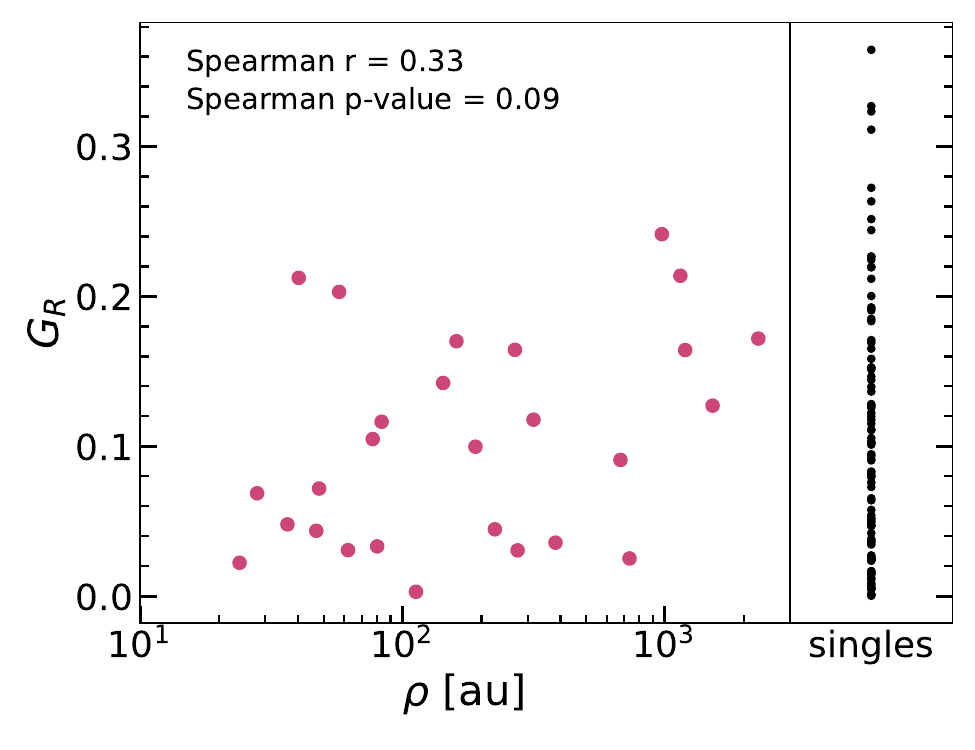}
    \caption{Planet radius Gini index $G_R$ vs binary projected physical separation $\rho$ (pink, left panel) with single stars plotted for comparison (black, right panel). The data are suggestive of a binary-separation dependent Gini index, where smaller-separation binary stars tend to have systems with lower Gini indices and thus more equal-sized planets. The two outliers in the upper left corner are unusual planetary systems, KOI-112 and KOI-750, and some outliers are expected because of our assumption that there are no split systems among the planetary systems around binary stars.}
    \label{fig:gini_sep}
\end{figure}

As discussed above, a binary star companion may shape the planetary system architectures of the planet host star via two main channels: (1) disrupting protoplanetary disks during the early stages of planet formation, and (2) dynamical post-processing of the young planets after the gas disk disperses. The first channel (protoplanetary disk disruption) should have a strong dependence on binary star separation because a closer binary will truncate the disk at a smaller radial distance and will provide greater XUV irradiation which may hasten the evaporation of the gas disk, shortening its lifetime. The second channel (post-formation dynamics) may have a more modest dependence on binary star separation in our sample, because the companions are typically at large separations compared to the planetary semimajor axes. Although the dynamical effects from a companion should also have some separation dependence (since the gravitational disruption from a companion increases with decreasing binary semimajor axis), we expect that these processes should be less sensitive to separation than processes which shape the protoplanetary disk, which can be altered by companions at much larger semimajor axes. This interpretation also follows from the results of \citet{Venturini2026}, who found that many of the observed properties of the population of planets in binary star systems can be explained via disk truncation without any additional dynamical perturbations.


Investigating planetary system architectures as a function of binary star separation may clarify the roles the above processes play during planet formation and evolution. Figure \ref{fig:gini_sep} shows the planet radius Gini index $G_R$ of each multi-planet system with a binary-star host plotted against binary projected physical separation $\rho$. By eye, it appears that there may be a relationship between $\rho$ and $G_R$. However, the statistical evidence for a correlation is not statistically significant, with a Spearman rank order coefficient calculated as $r_s = 0.33$ ($p = 0.09$). Rather than a strong correlation, we postulate that instead there may be an upper envelope on $G_R$ that decreases with decreasing binary separation. Comparison to the single-star planet population (Figure \ref{fig:gini_sep}, right panel), which may be thought of as an effective $\rho \rightarrow \infty$, supports the plausibility of this interpretation. The two outlier systems at low $\rho$, high $G_R$ are KOI-112 and KOI-750 which also exhibit unusual spacing architectures (see Figures \ref{fig:2pl_abacus} and \ref{fig:multi_abacus}, respectively); we speculate that these may in reality be ``split'' systems wherein some of the planets orbit the primary star and the rest orbit the secondary.

The tentative relationship between $\rho$ and $G_R$ aligns with the results of \citet{Sullivan2024}, who found that close binaries tend to host fewer sub-Neptunes than wider binaries. An upper envelope on $G_R$ might then be expected, as planets in close binary-star would tend to be more similar in size. This picture also agrees with theoretical understanding of planet formation. A close binary will more severely truncate the protoplanetary disk and inhibit mass accretion, thereby leading to a smaller maximum $G_R$ because mainly super-Earths are produced. Conversely, a wide binary does not prohibit the formation of systems of only super-Earths or only sub-Neptunes, but does allow mixed systems of both super-Earths and sub-Neptunes.

\begin{figure*}
    \includegraphics[width = 0.49\linewidth]{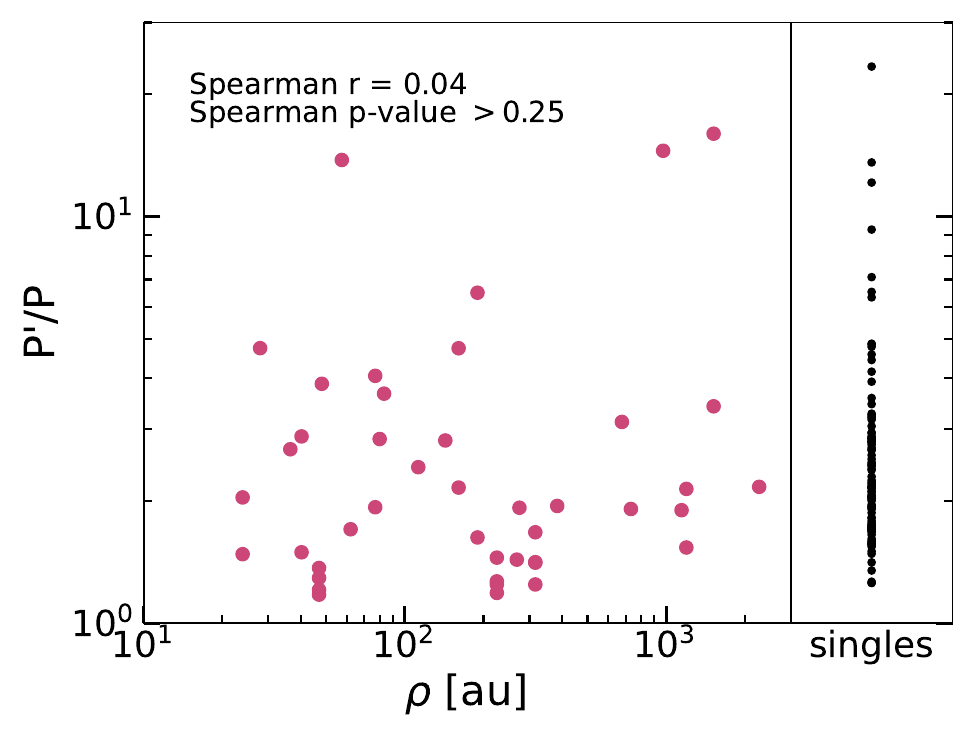}
    \includegraphics[width = 0.49\linewidth]{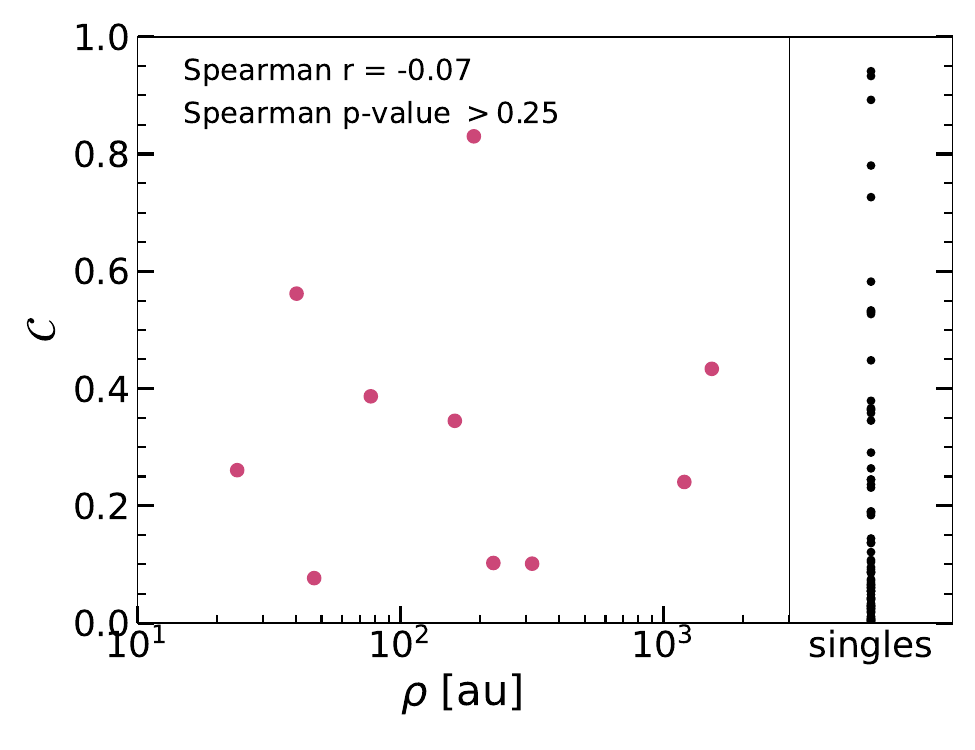}
    \caption{Period ratio (left) and gap complexity (right) plotted as a function of binary star projected physical separation, with single-star systems plotted for comparison on the right side of each figure. Neither quantity shows any apparent relationship between planet spacing and binary star separation.}
    \label{fig:sep_period_C}
\end{figure*}

Figure \ref{fig:sep_period_C} shows the relationship between binary separation and planet spacing architecture; we identify no strong or even tentative relationship between these quantities. We also previously found that planet multiplicity distribution of wide- and small-separation binary-star hosts is indistinguishable (Figure \ref{fig:multiplicity_fraction}), These non-detections of relationships between $\rho$ and $N$, $P'/P$, or $\mathcal{C}$ suggests that the dynamical evolution of planetary systems with binary star companions is only weakly dependent on the precise properties of the binary.

The higher typical gap complexity in binary-star systems compared to single-star systems is reminiscent of the tendency for compact multis with outer giant planet companions between ${\sim}1-10$ au to have higher gap complexities compared to those without outer giants \citep{HeWeiss2023}. It may be that any large external body to a compact multi induces dynamical excitation with similar results over a wide range of companion masses and separations. So far, theoretical models have not reached a consensus on whether dynamical perturbations from the external companion can induce mutual inclinations necessary to reproduce observed trends in $N$ and $\mathcal{C}$ \citep{LammersWinn2025, LiveseyBecker2025}.

Taken together, the tentative relationship between planet size architecture and binary separation juxtaposed against the lack of a relationship between planet spacing architecture and binary separation bolsters the interpretation that collisions and other violent dynamical processes do not dominate when planetary systems are forming around binary star hosts. Instead, it appears that the original properties of the disk (size, lifetime, mass) set planetary system architectures at formation, and secular dynamical processes are a second-order effect.

\section{Summary and Conclusions}\label{sec:conclusion}
We assembled a sample of 162 transiting planets in 118 binary-star systems from \citet{Sullivan2024} and compared them to a sample of 880 planets in 544 single-star systems from the California-Kepler Survey \citep{Johnson2017}. We assessed the planetary architectures of the systems, exploring planet radius self-similarity, planet period uniformity, and planet multiplicity to investigate whether planetary systems with binary-star hosts have substantially different architectures than those with single-star hosts.

We found no statistically significant differences between the planet size architectures of binary-star systems compared to the planet size architectures of single-star systems, as determined by assessing planet radius ratios $R_p'/R_p$ and planet radius Gini indices $G_R$. In contrast, we identified a $2.5\sigma$ difference in the planet spacing architectures, with binary-star planets exhibiting both tighter and more uneven spacing, as determined by assessing planet period ratios $P'/P$ and gap complexities $\mathcal{C}$. We further identified a $3 \sigma$ difference in planet multiplicity, with binary-star hosts showing a higher prevalence of single-planet systems.

We interpret these results as evidence that stellar binaries primarily influence planet formation by altering the properties of their protoplanetary disks. Planets in binary-star systems form in smaller, shorter-lived disks, which leads to systematically smaller planets (see, e.g., \citealt{Sullivan2024}) that can be more tightly packed in terms of period ratios while still maintaining Hill stability. A lower average planet multiplicity for binary-star hosts also follows naturally from the suppressed occurrence rate of planets in binary-star systems \citep[e.g.,][]{Moe2021, Sullivan2026}. In this picture, planetary system architecture is largely set at formation by the properties of the protoplanetary disk \citep[see also][]{Venturini2026}, with dynamical evolution continuing to sculpt planetary architectures over secular timescales after the protoplanetary disk dissipates. 

Future follow-up should prioritize Doppler measurements for the binary star systems to seek additional non-transiting planets and measure planet masses, which would clarify what role violent dynamical processes like ejections and collisions play during the formation and evolution of planetary systems with binary star hosts. We also intend to pursue follow-up work to measure planet orbital eccentricities from the photometric lightcurves and perform detailed N-body dynamical analyses of the planetary systems.

Overall, we identified relatively minor differences in the architectures of single-star planetary systems vs. binary-star planetary systems. This similarity suggests that the physics of (small) planet formation are largely unaltered by the presence of a second star in a planetary system except insofar as binary stars alter the properties of their stellar companion's protoplanetary disks. This conclusion cements the utility of binary star systems as natural laboratories for studying planet formation. With space-based surveys expected to identify hundreds of new binary stars and planets in the near future \citep{LammersWinn2026}, we eagerly anticipate the new insights that will be gleaned from studying this rich and fascinating population.

\begin{acknowledgments}
We thank the referee for their comments, which improved this paper. This work made use of the gaia-kepler.fun crossmatch database created by Megan Bedell. This work has made use of data from the European Space Agency (ESA) mission {\it Gaia} (\url{https://www.cosmos.esa.int/gaia}), processed by the {\it Gaia} Data Processing and Analysis Consortium (DPAC, \url{https://www.cosmos.esa.int/web/gaia/dpac/consortium}). Funding for the DPAC has been provided by national institutions, in particular the institutions participating in the {\it Gaia} Multilateral Agreement. This research has made use of the VizieR catalogue access tool, CDS, Strasbourg, France (DOI : 10.26093/cds/vizier). The original description  of the VizieR service was published in 2000, A\&AS 143, 23
\end{acknowledgments}

\bibliography{bib}
\bibliographystyle{aasjournal}

\restartappendixnumbering
\appendix

\section{Results of the analysis while implementing split systems}\label{sec:appendix_secondary_mix}
In the main body of the text, we assumed that all of the systems in the binary star sample were comprised exclusively of planets orbiting the primary star, and argued that this assumption is justified on both astrophysical and observational grounds. However, the assumption of primary-host-only may not be the case in every system. Thus, we also performed the analysis by replicating the assumption of \citet{Sullivan2024} that 20\% of the planets orbit the secondary star instead of the primary.

We implemented this assumption by randomly assigning 20\% of planets independently to secondary star hosts instead of primary star hosts. In this scenario some fraction of the multi-planet systems become ``split'', with some planets around the primary and some around the secondary. Because most multis are two-planet systems, the most common outcome when a split occurs is to convert a single two-planet system into two one-planet systems, thereby reducing the number of multi-planet systems in the sample and diminishing the statistical strength of the results. The planet multiplicity distribution is most strongly impacted, while radius ratios, Gini size indices, period ratios, and gap complexity are more indirectly affected.

\begin{figure*}
    \centering
    \includegraphics[width = 0.45\linewidth]{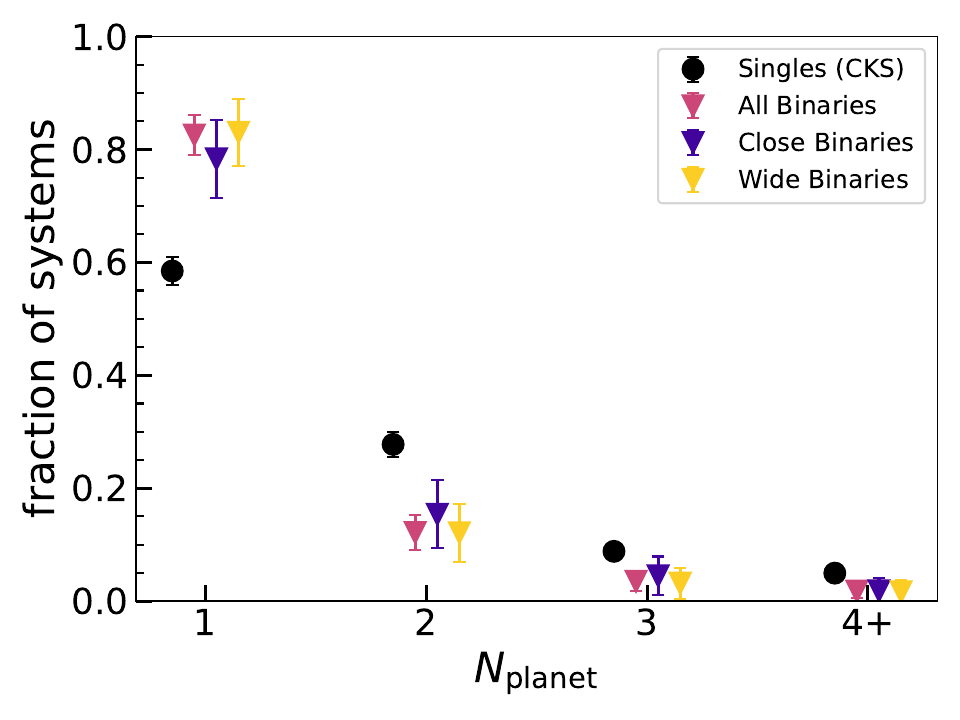}
    \caption{Top row: Planet multiplicity fractions for single stars (black), close binaries (blue), wide binaries (yellow) and all binaries (magenta), assuming split systems (80\% primary, 20\% secondary). The fraction of single-planet systems increases because of split systems (i.e., two-planet systems where each star hosts one planet would be counted as two single-star systems). The remainder of the multiplicity fractions remain largely unchanged from the original analysis.}
    \label{fig:sec_mix_multiplicity}
\end{figure*}

\begin{figure*}
    \centering
    \includegraphics[width = 0.9\linewidth]{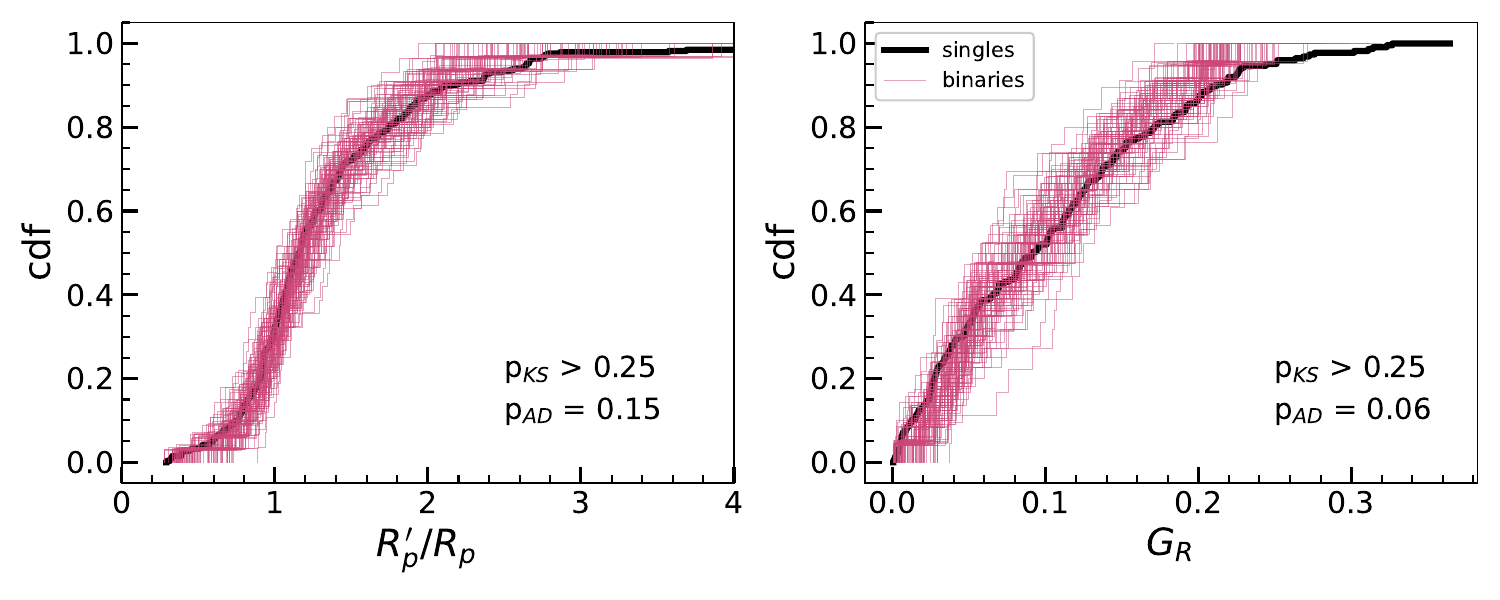}
    \caption{The cumulative distributions of radius ratios (left) and Gini indices (right; as for Figure \ref{fig:radius_ratios}) for a 20\% secondary star planet host mixture. The distributions are similar to the 100\% primary analysis, likely because the total number of multi-planet systems decreases and the number of split systems is relatively small.}
    \label{fig:sec_mix_radius}
\end{figure*}

\begin{figure*}
    \centering
    \includegraphics[width = 0.9\linewidth]{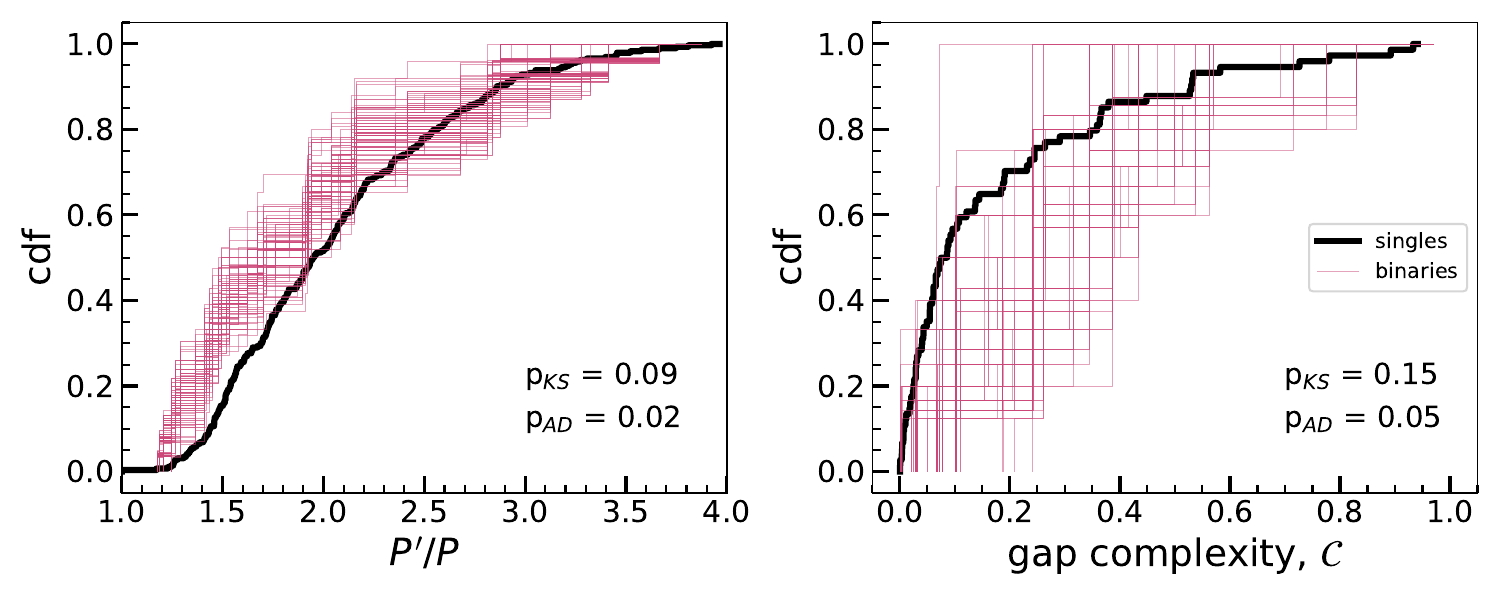}
    \caption{Period ratios and gap complexity for a 20\% secondary host star mixture. The statistical significance of differences in the period ratios and complexities decreases because of the smaller sample, but from visual inspection the differences between the single star and binary star distributions remain substantial.}
    \label{fig:sec_mix_complexity}
\end{figure*}

We bootstrapped each calculation 100 times, randomly selecting primary versus secondary star hosts for each iteration. The results from this analysis are shown in Figures \ref{fig:sec_mix_multiplicity}, \ref{fig:sec_mix_radius}, and \ref{fig:sec_mix_complexity}. In general, differences between the 100\% primary star and 80\% primary star samples are small, for the reasons discussed above. We conclude that our analysis is robust against the assumption of primary-only planets.

\section{Results of the analysis with a planetary radius cut}\label{sec:appendix_rp_cut}

We performed the analysis in the main body of the text without imposing a radius cut on the planetary systems, but \citet{Sullivan2024} and \citet{Sullivan2026} found that sensitivity to small planets ($R_{p} < 1.3 R_{\oplus}$) decreases rapidly in binary-star systems. Thus, we also tested our analysis with a radius cut imposed on both the binary and single-star systems, requiring that all planets have $R_{p} > 1.3 R_{\oplus}$. In general, we did not observe significant differences between the analyses with and without the radius cut, but the radius-cut figures are presented here for completeness and reference. Notably, the explanatory power of several tests is reduced because the 1.3 $R_{\oplus}$ cut reduces the number of high-order-multiple ($N \geq 3$) systems to 2, and the number of multi-planet systems to 16 total. The distributions of planet architecture measures for single-star systems do not change substantially from the radius cut, likely because the sample is large enough that removing some planets does not strongly impact the distributions.

\begin{figure*}
\centering
    \includegraphics[width=0.5\linewidth]{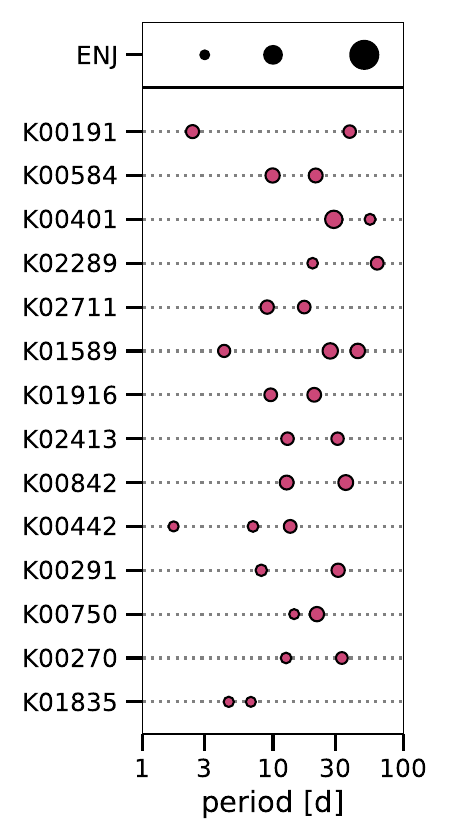}
    \caption{All multi-planet systems with binary-star hosts, after removing all planets with $R_p < 1.3 \Re$. Systems are ordered by binary separation, with close separations at the bottom and wide separations at the top. All $R_{p}$ assume the planets orbit the primary star. Compared to the analysis in the main text (Figure \ref{fig:2pl_abacus}, the number of systems with 2 planets has been reduced from 17 to 12, and the number of systems with 3+ planets has been reduced from 10 to 2.}
    \label{fig:app:abacus}
\end{figure*}

\begin{figure}
    \centering
    \includegraphics[width=0.5\linewidth]{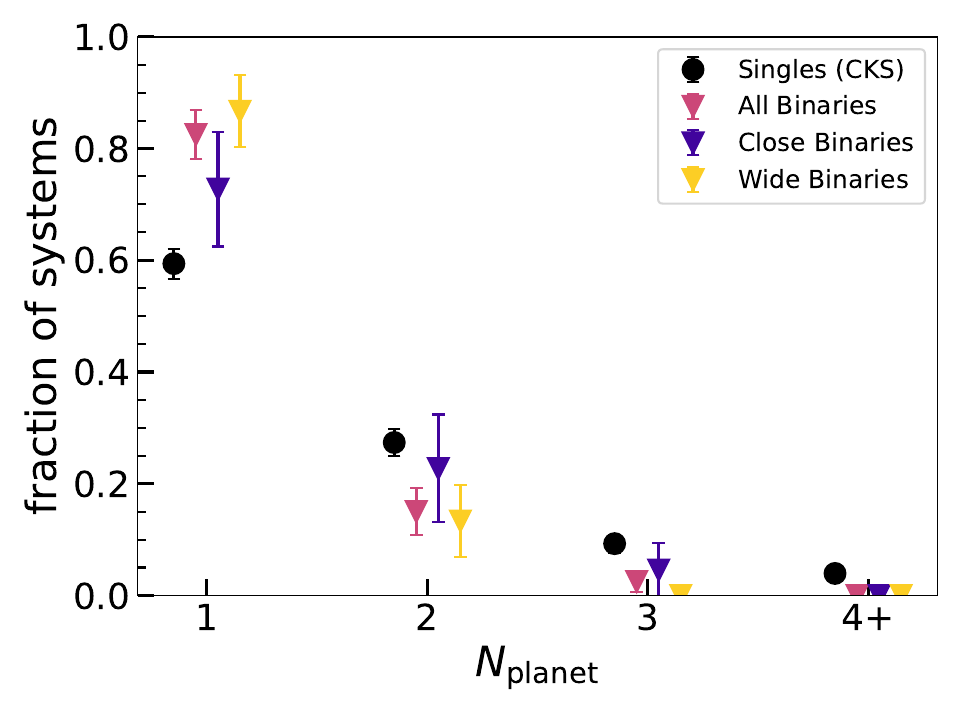}
    \caption{Fraction of planetary systems with 1, 2, 3, or 4+ planets detected for single stars (black circles), all binary stars (pink triangles), close binary stars (blue triangles) and wide binary stars (yellow triangles), after removing all planets with $R_{p} < 1.3$\,\Re. The fraction of single-planet systems is larger for all binary star planet host cases, and is statistically significant (3.0$\,\sigma$) based on a $\chi^{2}$ contingency test for all stellar binaries versus stellar singles.}
    \label{fig:app:multiplicity_fraction}
\end{figure}

\begin{figure}
    \centering
    \includegraphics[width=\linewidth]{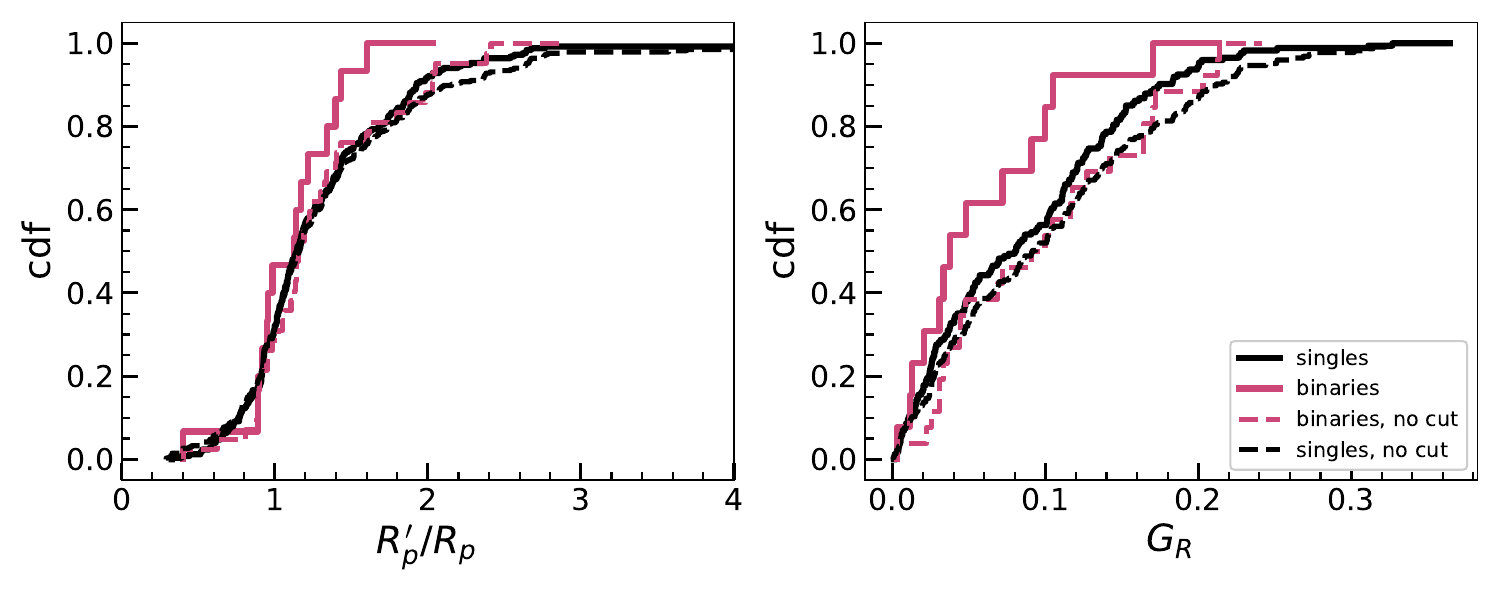}
    \caption{\textit{Left}: Cumulative distributions of the outer/inner radius ratios $R_p'/R_p$ for adjacent planet pairs in binary-star systems (pink) and single-star systems (black), after removing all planets with $R_p  < 1.3 \Re$. \textit{Right}: Cumulative distributions of planet radius Gini indices $G_R$. Compared to the analysis in the main text (Figure \ref{fig:radius_ratios}, values of $R_p'/R$ and $G_R$ have been shifted to lower values as a consequence of the now-limited range of detectable $R_p$. The p-values for the KS and AD tests between the planets in single stars and those in binary star systems for $R_p'/R_p$ and $G_R$ are $> 0.25$ in all cases, indicating that the apparent differences are not statistically significant, likely because of the greatly reduced sample size for the planets in binary stars. }
    \label{fig:app:radius_gini}
\end{figure}

\begin{figure}
    \centering
    \includegraphics[width=0.5\linewidth]{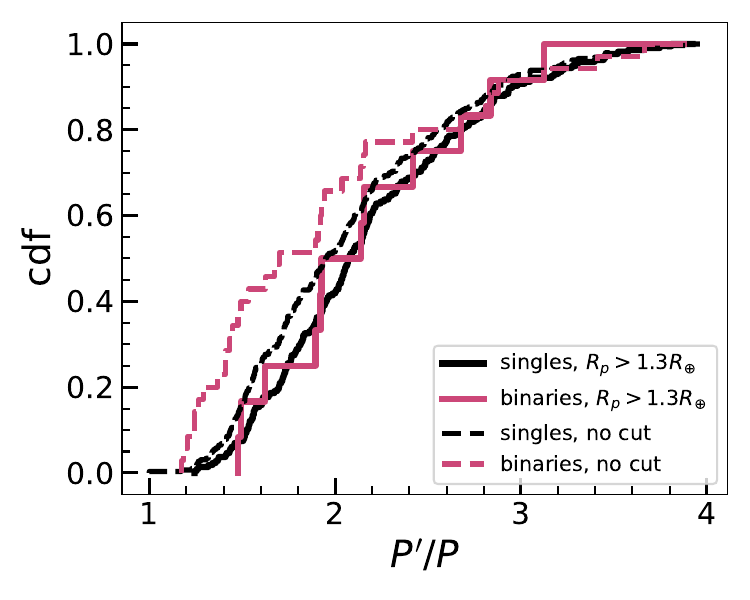}
    \caption{Cumulative distributions of the outer/inner period ratios $P'/P$ for adjacent pairs of planets in binary-star (pink) and single-star (black) systems, after removing all planets with $R_p  < 1.3 \Re$. The two distributions are now statistically indistinguishable, with KS and AD test p-values $>$ 0.25.}
    \label{fig:app:period_complexity}
\end{figure}

\end{document}